# Spectral fluctuations of tridiagonal random matrices from the $\beta$-Hermite ensemble

C. Male [a], G. Le Caër [b] and R. Delannay

Groupe Matière Condensée et Matériaux, C.N.R.S. U.M.R. 6626, Université de Rennes-I, Campus de Beaulieu, Bât. 11A, Avenue du Général Leclerc, F-35042 Rennes Cedex, France

[a] permanent address : Ecole Normale Supérieure de Cachan, Campus de Kerlann, F-35170 Bruz, France

[b] corresponding author, E-mail : gerard.le-caer@ univ-rennes1.fr

## Abstract

A 'time series' $\delta_n$, the fluctuation of the $n$th unfolded eigenvalue, where $n$ plays the role of a discrete time was recently characterized for the classical Gaussian ensembles of $N \times N$ random matrices (GOE, GUE, GSE). It is investigated here for the $\beta$-Hermite ensemble as a function of the reciprocal of the temperature $\beta$ by Monte Carlo simulations. The ensemble-averaged fluctuation $\langle \delta_n^2 \rangle$ and the autocorrelation function vary logarithmically with $n$ for any $\beta > 0$ ($1 \ll n \ll N$). The simple logarithmic behavior reported in the literature for the higher-order moments of $\delta_n$ for the GOE $(\beta = 1)$ and the GUE $(\beta = 2)$ is valid for any $\beta > 0$ and is accounted for by Gaussian distributions whose variances depend linearly on $\ln n$. The $1/f^\alpha$ noise displayed by the $\delta_n$ series, previously demonstrated for the three Gaussian ensembles, is characterized by wavelet analysis both as a function of $\beta$ and of $N$. When $\beta$ decreases from 1 to 0, for a given and large enough $N$, the evolution from a $1/f$ noise at $\beta = 1$ to a $1/f^2$ noise at $\beta = 0$ is heterogeneous with a $\sim 1/f^2$ noise at the finest scales and a $\sim 1/f$ noise at the coarsest ones. The range of scales in which a $\sim 1/f^2$ noise predominates, grows progressively when $\beta$ decreases. Asymptotically, a $1/f^2$ noise is found for $\beta = 0$ while a $1/f$ noise is the rule for $\beta > 0$.

## I. INTRODUCTION

Random matrix theory (RMT) contributes significantly to quantum chaology which pertains to the statistical properties of quantum systems whose classical counterparts are chaotic [1-10]. The working definition of dynamical chaos for infinite quantum systems refers indeed to RMT [6,10]. As recalled by Prosen [10], a many-body quantum system is said to be chaotic if its excitation spectrum or some other dynamical characteristics are well described by those of ensembles of Hermitian random of appropriate symmetries on certain energy or time scales. The level fluctuations of a time-reversal symmetric quantum system were conjectured to coincide with those of the GOE for systems whose classical limit is chaotic [3]. In the semiclassical limit, the fluctuations of the energy levels of generic quantum systems, relative to their smoothed level densities, coincide in fact with those of eigenvalues of ensembles of random matrices chosen according to the physical symmetries of the considered systems. The converse is however not necessarily always true as, for instance, the classical counterparts of quantum systems showing GOE fluctuations may be regular [11].

The local spectral fluctuations of properly rescaled and processed eigenvalues of random matrix ensembles define universality classes in the limit of large matrix sizes which depend on the matrix symmetries and are independent on the details of the probability distributions of matrix elements. Such universality classes are for instance associated with the three fundamental Gaussian ensembles where $N \times N$ matrices are real symmetric for the Gaussian orthogonal ensemble (GOE), Hermitian for the Gaussian unitary ensemble (GUE) and quaternion self-dual for the Gaussian symplectic ensemble (GSE). A fourth ensemble, the Gaussian diagonal ensemble (GDE), is made from matrices whose sole non-zero elements are diagonal with identical and independent normal distributions.

An ubiquitous characteristic of short-range correlations is the asymptotic distribution of the spacing '$s$' between consecutive energy levels of quantum systems or between successive eigenvalues of random matrices, once unfolded [1-2,6,12]. Theoretical nearest-neighbor spacing (NNS) distributions are rarely available, simulated distributions are used instead and compared to exact or to approximate distributions of the reference ensembles. The Gaussian ensembles define for instance three universality classes of level repulsion at small '$s$', $s \to 0$, $p_{W,\beta}(s) \sim s^{\beta}$ with $\beta$=1, 2, 4 for the GOE, the GUE and the GSE respectively. The properties of eigenvalues of Gaussian ensembles are recalled to be the equilibrium characteristics at a temperature $1/\beta$ of $N$ identical point charges on a line in 2D which interact via a logarithmic Coulomb potential and are confined by an external harmonic potential [1]. The unfolded eigenvalues of a GDE matrix are independent and uniformly distributed. The asymptotic distribution of their spacings is thus a Poisson distribution $p(s) = \exp(-s)$ ($\langle s \rangle = 1$). Most often, phenomelogical models of the evolution of the NNS distributions are used to describe specific transitions between the Wigner-Dyson and the Poisson statistics.

Other classical characteristics of spectral fluctuations are the number variance and the spacing variance. The number variance, measures the $L$-dependence of the fluctuation of the number of unfolded eigenvalues in an interval of fixed length $L$ thrown at random on the eigenvalue sequence. By contrast, the spacing variance, measures the $n$-dependence of the fluctuation of the total length of a fixed number $n$ of spacings between successive unfolded eigenvalues. Both are simply related ([12-14] and section V below). The Dyson-Mehta statistic yields information about the spectral rigidity and long-range correlations by quantifying an average deviation of the cumulative level density from a line.

A different statistic, closely related to the level density fluctuation, was recently considered in a series of papers [12,15-27]. Named $\delta_n$ statistic, it is defined as :

$$\delta_n = \sum_{i=1}^{n}(s_i - 1) = \varepsilon_{n+1} - \varepsilon_1 - n \qquad (1)$$

where the spacing between two successive unfolded levels $\varepsilon_i$ and $\varepsilon_{i+1}$ is $s_i = \varepsilon_{i+1} - \varepsilon_i$. The fluctuation of the $n$th excited state, $\delta_n$, is then considered as a time series of size $M$ where $n$ plays the role of a discrete time. The power spectra were shown to display a $1/f^\alpha$ power law behaviour, where $f = 2\pi k/M$ ($k = 1,..,M$) is the frequency, with an exponent $\alpha$ of 2 for the GDE and of 1 for the GOE , the GUE and the GSE. Relaño et al. [15] conjectured then that the energy spectra of chaotic quantum systems are characterized by $1/f$ noise. That result was subsequently derived by Faleiro et al. from random matrix theory with a power spectrum $S(k)$ varying as $N/\beta k$ for chaotic systems and as $N^2/k^2$ for integrable systems when the matrix size $N$ is large and $k \ll N$ [16]. Computer calculations of the energy level fluctuations of various chaotic quantum systems indicated that the behavior $1/f$ is robust.

Various authors tried to benefit from the simplicity and from the numerical efficiency of computer simulations of ensembles of tridiagonal random matrices with one control parameter to model chaotic quantum systems. In that way, the level-spacing statistics of the model can be changed from GOE-like to Poisson-like [20,28]. Molina et al. [20] used a Lanczos algorithm to reduce a Hamiltonian matrix to a tridiagonal form and established an approximate relation between the control parameter of the model and $\beta$. The $\beta$-Hermite ensemble ($\beta$-HE) [29-30], whose fluctuations characteristics were recently studied [30-32], is an ensemble of real-symmetric tridiagonal matrices family for which the the temperature $1/\beta$ can be chosen at will. All its spectral characteristics coincide with those of the three classical Gaussian ensembles for the corresponding values of $\beta$ ($\beta$ =1,2,4). The aim of the present paper is then to investigate the $\delta_n$ statistic in the $\beta$-Hermite ensemble as a function of $\beta$.

## II. THE $\beta$–HERMITE ENSEMBLE [29-30]

Random matrices from the $\beta$-Hermite ensemble are real-symmetric and tridiagonal. The joint distribution of the eigenvalues $(\lambda_1, \lambda_2,..., \lambda_N)$ of $N \times N$ $\beta$-Hermite matrices is, whatever $\beta \geq 0$ [29]:

$$\begin{cases} P_{N,\beta}(\lambda_1,..,\lambda_N) = K_{N,\beta} \exp\left(-\frac{1}{2\sigma^2}\left[\sum_{k=1}^{N}\lambda_k^2\right]\right)\left[\prod_{1 \leq j < k \leq N}|\lambda_j - \lambda_k|^\beta\right] \\ \rho = \frac{\beta}{2} \quad N_\rho = N + \rho N(N-1) \quad K_{N,\beta} = \sigma^{-N_\rho/2}(2\pi)^{-N/2}\prod_{j=1}^{N}\frac{\Gamma(1+\rho)}{\Gamma(1+j\rho)} \end{cases} \quad (2)$$

where $K_{N,\beta}$ is the reciprocal of the Mehta integral ([1] p.354). The eigenvalue distribution (eq. 2) is, for $\beta = 0, 1, 2, 4$, identical with those of the GDE, the GOE, the GUE and the GSE respectively. A $N \times N$ random matrix from the $\beta$-HE is defined as:

$$\mathbf{A}_{N,\beta} = \sigma \mathbf{H}_{N,\beta} = \sigma \begin{bmatrix} H_{11} & H_{12}/\sqrt{2} & 0 & . & 0 \\ H_{12}/\sqrt{2} & H_{22} & H_{23}/\sqrt{2} & 0 & . \\ 0 & H_{23}/\sqrt{2} & . & . & 0 \\ . & 0 & . & H_{N-1,N-1} & H_{N-1,N}/\sqrt{2} \\ 0 & . & 0 & H_{N-1,N}/\sqrt{2} & H_{NN} \end{bmatrix} \quad (3)$$

where $\sigma$ is a scale factor. The $2N-1$ distinct matrix elements are independent random variables. The diagonal elements are identically and independently distributed (iid). The $H_{kk}$'s $(k = 1, ..., N)$ are standard normal random variables $N(0,1)$. The off-diagonal element $H_{k,k+1}$ $(k = 1, ..., N-1)$ has a chi distribution with $k\beta$ degrees of freedom whose probability density is:

$$q_{N,\beta}(x) = 2^{1-k\beta/2} x^{k\beta-1} \exp(-x^2/2)/\Gamma(k\beta/2) \qquad (x \geq 0) \qquad (4)$$

When $\beta \to \infty$, $H_{k,k+1}$ can be written as $\sqrt{k\beta} + X/\sqrt{2}$, where $X$ is a standard Gaussian [29-30]. Then, the properly rescaled eigenvalues, whose large $\beta$ distribution tends to a multivariate Gaussian distribution, fluctuate around the $N$ roots of the $N$th Hermite polynomial [29-30,34].

### III. COMPUTER SIMULATIONS

We performed Monte Carlo calculations of random matrices from the $\beta$-Hermite ensemble both in Fortran and in Matlab with standard laptop computers. Gaussian variables were generated by the polar Box-Muller method [35]. The chi distributions of the non-diagonal elements were generated through gamma distributions ([35] p.410 and 418). The scale parameter (eq. 3), $\sigma = 1/\sqrt{4 + 2\beta(N-1)}$, was chosen so that $\langle \lambda^2 \rangle = 1/4$. In that way, the asymptotic eigenvalue distribution for $\beta > 0$ is a Wigner semi-circle of radius 1:

$$\rho_W(\lambda) = \frac{2}{\pi}\sqrt{(1-\lambda^2)} \qquad (5)$$

for $|\lambda| \leq 1$ and 0 elsewhere. The eigenvalue density is, to an excellent approximation, a Wigner semi-circle even for moderate values of $N$ (some tens) when $\beta$ ranges between ~1 and ~5 while deviations occur both for low and for high values of $\beta$ (figure 1 and [32]). The density at high temperature (small values of $\beta$) evolves from a smooth shape intermediate between that of a Gaussian and that of a Wigner semi-circle to a Wigner semi-circle when the matrix size increases (figures of [32]). At low temperature, the progressive freezing of charges around their equilibrium positions produces oscillations of the eigenvalue density around the smooth Wigner semi-circle [30,36] (figure 1). The appearance of the density in figure 1 for $\beta = 16, 32$ is strongly dependent on the bin size as it results from an interplay between the local wavelength of the previous oscillations and the bin size which is here 0.01. The matrix size must be significantly increased to damp such oscillations for a fixed bin size.

Once generated and diagonalized, the eigenvalue spectrum of a $N \times N$ $\beta$-Hermite matrix is 'unfolded' to calculate various spectral fluctuations. Any eigenvalue $\lambda_k$, which belongs to the interval $(-r, +r)$, with typical values of $r\,(<1)$ ranging between 0.8 and 0.9, is transformed into an unfolded eigenvalue $\lambda_k^{(u)}$. The latter is the value for $\lambda = \lambda_k$ of the cumulative distribution function, $F(\lambda) = \int_{-\infty}^{\lambda} \rho(x)dx$ of the smoothed level density $\rho(x)$ which is either calculated exactly or estimated numerically. As stressed by Gómez et al. [37], a correct unfolding procedure is needed to extract trustworthy characteristics of fluctuations, in particular for systems whose mean level density is unknown. Misleading results might be obtained when the latter density is replaced by a local mean level density calculated from a small number of levels around the level to unfold. The asymptotic level density of the $\beta$-Hermite ensemble is known to be a Wigner semi-circle (eq. 5) which can be reached for reasonable values of $N$ in a sufficiently broad range of $\beta$ to avoid the latter difficulties. When the eigenvalue density differs negligibly from a Wigner semi-circle in the selected range, the unfolding process is then performed as follows:

$$\lambda_k^{(u)} = \int_{-1}^{\lambda_k} \rho_W(\lambda)d\lambda = \frac{1}{2} + \frac{\lambda_k\sqrt{1-\lambda_k^2}}{\pi} + \frac{\sin^{-1}\lambda_k}{\pi} \qquad (6)$$

The unfolded density of eigenvalues is constant by construction. The unfolded eigenvalues are further rescaled so that the average spacing between nearest neighbors is $\langle s \rangle = 1$. When the empirical cumulative distribution shows significant deviations, mostly global for $\beta < 1$ [32], from a Wigner semicircle, the unfolding process was performed from a smooth eigenvalue density obtained numerically as the average of an ensemble of spectra simulated with Matlab. The simulated distributions and the 'time series' investigated in the present paper were obtained altogether from simulations of $10^6$ matrices with $N$=25, of $10^5$ matrices with $N$=200 and of $2.10^3$ matrices with $N$=513, 1000 matrices with $N$=2049, 500 matrices with $N$=4097, 500 matrices with $N$=8193 and 50 matrices with $N$=32769.

### IV. THE NNS DISTRIBUTIONS

One of the very widespread characteristic of fluctuations of eigenvalues of random matrices is the asymptotic distribution of the nearest-neighbor spacing (NNS) '$s$' between successive unfolded eigenvalues. NNS distributions of the $\beta$–HE are discussed in detail in [32] where generalized gamma (GG) distributions are shown to be excellent approximations of the simulated distributions for any $\beta$. A generalized gamma distribution, whose general form is:

$$p_{\omega_1,\omega_2}(s) \propto s^{\omega_1} \times \exp(-cs^{\omega_2}) \tag{7}$$

depends on two shape parameters $\omega_1$ and $\omega_2 = 2 - \omega$. The best least-squares approximation of the NNS distribution of the $\beta$-HE has $\omega_1 = \beta$ and a deviation to 2, $\omega$, which is well approximated by a stretched exponential [32]. It reads:

$$\begin{cases} p_{\beta,\omega}(s) = \left[(2-\omega)c_2/c_1^2\right](\alpha_{\beta,\omega}s)^{\beta} \exp\left(-(\alpha_{\beta,\omega}s)^{2-\omega}\right) \\ c_n = \Gamma((n+\beta)/(2-\omega)) \qquad \alpha_{\beta,\omega} = c_2/c_1 \qquad \omega = \exp(-2.12\beta^{0.75}) \end{cases} \tag{8}$$

as $\langle s \rangle = 1$. The level repulsion for small spacings varies thus as $s^{\beta}$, whatever $\beta > 0$, as expected from the Wigner surmise (eq.7 with $\omega_1 = \beta$ and $\omega_2 = 2$) [32]. For $\beta > \sim 2$, the best approximate distribution (eqs 7 and 8) differs little from the Wigner surmise. When $\beta \to \infty$, the expansion of the coefficient $\alpha_{\beta,0}$ of the Wigner surmise is:

$$\alpha_{\beta \to \infty, 0} = \frac{\beta}{2} + \frac{1}{4} + \frac{1}{16\beta} + O\left(\frac{1}{\beta^2}\right) \tag{9}$$

while that of $2\alpha_{\beta,0}^{\beta}c_2/c_1^2$ reads:

$$2\alpha_{\beta,0}^{\beta} c_2 / c_1^2 = \frac{\exp(\beta/2)}{\sqrt{\pi}}\left\{\sqrt{\beta} + \frac{1}{3\sqrt{\beta}} + O\left(\frac{1}{\beta\sqrt{\beta}}\right)\right\} \quad (10)$$

Then, $y = (s-1)\sqrt{2\beta}$ has an asymptotic standard normal distribution when '$s$' has a Wigner distribution (see too figure 8 of [32]), a result of interest for section V.2.

## V. THE $\delta_n$ STATISTIC OF THE $\beta$-HERMITE ENSEMBLE

A finite series interpreted as a discrete 'time' series was investigated recently [15-19]. It is defined as $\delta_n = \sum_{i=1}^{n} s_i - n$, $(n = 1, 2, ..., N)$ (eq. 1) and was shown to display $1/f$ noise for the three classical Gaussian ensembles. Fluctuations of $\delta_n$ were considered earlier by Brody et al. (eq. 5.5 of [12]). Relaño et al. [19] calculated the spacing correlation functions, $C_{q,\beta}(n)$, for the GOE and the GUE:

$$C_{q,\beta}(n) = \left(\left|\delta_{m+n} - \delta_m\right|^q\right)_m = \frac{1}{N'}\sum_{m=1}^{N'}\left|\delta_{m+n} - \delta_m\right|^q \quad (11)$$

with:

$$\delta_{m+n} - \delta_m = \sum_{i=m+1}^{m+n} s_i - n = S_n(m) - n \quad (12)$$

where $1 \ll N' \leq N - n$ is the number of points of the given realization over which the 'time' or spectral average, denoted as $(..)_m$, is taken. To calculate $C_{q,\beta}(n)$, Relaño et al. [19] performed actually a double average, namely a spectral average followed by an ensemble average. A few $\delta_n$ series are shown in figure 2.

### V.1. The fluctutation and the autocorrelation
#### V.1.1. General characteristics

The spacing variance $\sigma_\beta^2(n) = \left\langle\left(\delta_{m+n} - \delta_n\right)^2\right\rangle$ was defined as the variance of the sum of a fixed number $n$ of consecutive nearest-neighbor spacings (eq. 12) [12, 14]. By stationarity, that variance was concluded to be independent on $m$ (eq.12). For a Poisson process, $\beta = 0$, the spacing variance is simply $\sigma_0^2(n) = n$. By contrast, the number variance, denoted here as $\Sigma_\beta^2(L)$, measures the fluctuation of the number of unfolded eigenvalues in an interval of fixed length $L$ thrown at random on the considered sequence of eigenvalues. It is known for the Gaussian ensembles in the limit of large $L$ [1,14]:

$$\Sigma_\beta^2(L) = \frac{2}{\beta\pi^2}\left(\ln L + \gamma + \ln(2\pi) + 1\right) + c(\beta) + O\left(\frac{1}{L}\right) \quad (13)$$

where $\gamma = 0.5772215...$ is the Euler constant, $c(1) = -1/4$, $c(2) = 0$ and $c(4) = \ln 2/(2\pi^2) + 1/16$ while $\Sigma_0^2(L) = L$. For many ensembles including the Gaussian ensembles [12-14], the number variance evaluated at $L=n$ and the spacing variance are, to good precision, related for large $n$ by:

$$\sigma_\beta^2(n) = \Sigma_\beta^2(n) - 1/6 \quad (14)$$

The constant $-1/6$ must be removed for a Poisson process. By expanding the logarithm of $P_{N,\beta}(\lambda_1,..,\lambda_N)$ (eq.2 with $\sigma^2 = 1/\beta N$) in the vicinity of its maximum, Andersen et al. [34] used the resulting multivariate Gaussian distribution of the eigenvalues (see also [30]) to derive the small amplitude normal modes of the spectrum describing the fluctuations of the eigenvalues around their equilibrium positions. They showed that the $k$th unfolded eigenvalue of the central part of the spectrum can be approximated for large $N$ by [34]:

$$x_k = \frac{\pi k}{N\sqrt{2}} + \frac{1}{\sqrt{N}}\sum_{j=1}^{N}\alpha_j U_{j-1}\left(\frac{\pi k}{2N}\right) \quad (15)$$

with the scaling $\sigma = 1/\sqrt{\beta N}$ (eq. 2). The $U_{j-1}(x)$'s are Chebyshev polynomials of the second kind and the $\alpha_j$'s are independent $N(0, 1/j\beta N)$ Gaussians. They derived from eq. 15 the logarithmic term of the number variance (eqs 13 and 14), valid whatever $\beta$ for large $L$. Eq. 15 can be used to derive directly the logarithmic term of $\sigma_\beta^2(n)$ as shown in appendix A. The $\ln n$ ($r.s.p.$ $\ln L$) dependence of the spacing (r.s.p. number) variance stems actually from the variances $1/j$ of the $\alpha_j$'s (eq. 15 and appendix A). The exact lower order terms are not obtained by the approximation method which yields eq. 15 [34].

**V.1.2. Numerical simulations**

We show in figure 3 the ensemble-averaged variance $\sigma_1^2(n)$ for several values of $N$ for $\beta = 1$ in the whole range of unfolded eigenvalues ($N' = 2^r$ values (eq.11) for $N = 2^{r+1} + 1$). It increases for small values of $n$ as $a_\beta \ln n + cst$ and then flattens with a remaining dependence on $n$ which is symmetric with respect to the center of the considered range.

As suggested both by eq. 15 (appendix A) and by the times series considered in the next section (see eq. 33 below), the overall shape of $\sigma_1^2(n)$ was tentatively least-squares fitted by $\sigma_1^2(n) = a_1 \ln(\sin^2(\pi n/N')) + b_{1,N} + c_{1,N}(1-2n/N')^2$ for values of $N$ ranging between 129 and 8193. They are in fair agreement with the simulated curves as shown by figure 3. As expected from eq. 13, $\beta a_\beta = 0.103 \pm 0.007 \simeq 1/\pi^2$ and gives as above a $2\ln n/\beta\pi^2$ term when $1 \ll n \ll N$ ($n \leq n_M \sim 0.1N$). The rather flat central region, $n \sim N/4$, is seen to rise logarithmically with $N$ (figure 3).

The behavior for $1 \ll n \ll N$ was more thoroughly investigated. Numerical values of $\sigma_\beta^2(n)$ were calculated by spectral averaging followed by ensemble averaging from the central parts of sequences of unfolded eigenvalues obtained from Monte-Carlo simulations of five hundred $4096 \times 4096$ $\beta$-Hermite matrices (figure 4). Writing then:

$$\sigma_\beta^2(n) = A_{2,\beta} \ln n + B_{2,\beta} \qquad (16)$$

we calculated the coefficients $A_{2,\beta}$ and $B_{2,\beta}$ shown in figure 5. The parameter $\beta A_{2,\beta}$ is independent of $\beta$ with an average value of 0.205(4), consistent again with that of $\frac{2}{\pi^2} = 0.202642...$ The second parameter $\beta B_{2,\beta}$ increases first rapidly for small $\beta$ and then slowly (figure 5). From eq.16, if we ignore that the latter holds only for large $n$, we deduce that $\sigma_\beta^2(1) = B_{2,\beta}$. Actually, figure 1 of Relaño et al. [19] shows that eq.16 is obeyed very precisely down to $n=1$. The variance of the GG distribution discussed in the previous section (eq. 8) reads:

$$\sigma_\beta^2(1) = \Gamma\left(\frac{1+\beta}{2-\omega}\right)\Gamma\left(\frac{3+\beta}{2-\omega}\right) \bigg/ \left[\Gamma\left(\frac{2+\beta}{2-\omega}\right)\right]^2 - 1 \qquad (17)$$

The dotted line in figure 5 shows $\beta\sigma_\beta^2(1)$ as a function of $\beta$. It accounts fairly well for the $\beta$ dependence of $B_{2,\beta}$, particularly for $\beta \geq 2$. The deviations between the exact values $\sigma_\beta^2(1)$ for the three Gaussian ensembles [33] and $B_{2,\beta}$, calculated from eqs 13 and 14 and eq. 16, decrease rapidly with $\beta$ (table 1). The spectrum of the unfolded zeros of the Hermite polynomials of large order, whose asymptotic distribution is a Wigner semi-circle too [38], becomes closer and closer to a rigid picket-fence spectrum at lower and lower temperature ([34,39], eq. 18 below). Eigenvalues and thus NN spacings '$s$' have Gaussian fluctuations around these zeros. The expansion for large $\beta$ of the variance $\sigma_\beta^2(1) = \langle (s-1)^2 \rangle_\beta$ is [32]:

$$\sigma_\beta^2(1) = \frac{1}{2\beta} - \frac{3}{8\beta^2} + ... \tag{18}$$

(eqs 9 and 10). Finally, we propose consistently that:

$$\lim_{\beta \to \infty} \{\beta \sigma_\beta^2(n)\} = \frac{2}{\pi^2} \ln n + \frac{1}{2} \tag{19}$$

in agreement with the trend seen in figure 5.

We investigated the autocorrelation function $K_\beta(n) = \langle (\delta_{m+n} \delta_m)_m \rangle$ for series constructed from eigenvalues of $N \times N$ $\beta$-Hermite matrices with $N = 2049$ $(0.25 \leq \beta \leq 8)$ and $N = 8193$. Half of the spectrum centered on 0 was unfolded and $\delta_n$ was constructed for $1 \leq n \leq (N-1)/8$, then $(\delta_{m+n} \delta_m)_m$, $1 \leq m \leq 3(N-1)/8$, was calculated and the ensemble average $K_\beta(n)$ was finally obtained (figure 6). The autocorrelation function varies logarithmically, $K_\beta(n) = -a_\beta \ln(n/N) + e_{\beta,N}$, with $\beta a_\beta = 0.100 \pm 0.002$ (appendix A) for $\sim 0.005 < n/N < \sim 0.1$ ($N = 2049, 8193$). A second method was used for $\beta = 1$ and different values of $N$ (figure 6). For a given value of $n$, the moving average was performed over the largest possible range of $m = 1,..,M(n)$. A slope identical with the previous one is found for small values of $n/N$ (figure 6), for larger values the autocorrelation function decreases almost linearly before reaching its minimum, close to zero, at $\approx 0.5$. Figure 6 suggests further that the overall level depends linearly on $\ln N$.

### V.2. Higher-order moments
### V.2.1. The $\beta$-Hermite ensemble

From their precise numerical simulations $(1 \leq q \leq 10)$, Relaño et al. [19] concluded that:

$$C_{q,\beta}(n) = \langle (|\delta_{m+n} - \delta_m|^q)_m \rangle = (A_{q,\beta} \ln n + B_{q,\beta})^{q/2} \tag{20}$$

Although it must be used with care for the considered time series, eq. 15 suggests that it tends to a discrete and wide-sense stationary Gaussian process [34, 39], whose increments are by definition stationary, as $\delta_{m+n} - \delta_m$ has a Gauss distribution with a zero mean and a variance, $\left\langle \left(\delta_{m+n} - \delta_m\right)_m^2 \right\rangle = \frac{2}{\beta\pi^2} \ln n + cst$. Indeed, $\delta_{m+n} - \delta_m$, which is a sum of a large number $n$ of spacings, is concluded to have a Gaussian distribution with a variance $\sigma_\beta^2(n)$ given by eq.16 as two $n$th order spacings $S_n(m_1) - n$ and $S_n(m_2) - n$ are uncorrelated when $m = |m_1 - m_2| > \approx 4-5$ (see figure 12 of [32]). When $n$ is large enough, a central limit theorem for $m$-dependent sequences applies to the considered spacings as described by Brody et al. in appendix N of [12] and in full agreement with our computer simulations. Such a Gaussian distribution gives:

$$C_{q,\beta}(n) = c_q \left(A_{2,\beta} \ln n + B_{2,\beta}\right)^{q/2} \qquad (21)$$

When $q$ has any positive real value, the moments of a standard Gaussian $N(0,1)$ yield $c_q = \left\langle |x|^q \right\rangle = \frac{2^{q/2} \Gamma\left((q+1)/2\right)}{\sqrt{\pi}}$. It comes then:

$$\beta A_{q,\beta} = \frac{4\left(\Gamma\left((q+1)/2\right)\right)^{2/q}}{\pi^{2+1/q}} = a_q \qquad (22)$$

The coefficient $a_q$ is independent of $\beta$. The previous conclusions are in agreement with the numerical simulations of Relaño et al. [19] for $q$ ranging between 1 and 10. They concluded from the latter that $A_{q,\beta}$ ($\beta = 1, 2$) is an exact linear function of $q$. From eq. 22, the increase of $a_q$ with $q$ is in fact almost perfectly, though not exactly, linear. Indeed, the Stirling formula applied to $c_q$ gives $c_q^{2/q} \approx q/e$ for large integer values of $q$. For $1 \leq q \leq 10$, the average slope is for instance $0.0742946..$ while its asymptotic value is $\left(da_q/dq\right)_\infty = 2/e\pi^2 = 0.0745479...$ Moreover, the values $a_2 = 0.20264..$ and $a_{10} = 0.79765..$ (eq. 22) and $A_{q,1} = 2A_{q,2}$ all agree with the results shown in figure 3 of [19]. We read on the latter that $A_{2,1} = a_2 \approx 0.2$ and $A_{10,1} = a_{10} \approx 0.8$ while we simulated $a_2 = 0.21(1)$. For large $\beta$, the asymptotic standard normal distribution of $y = (s-1)\sqrt{2\beta}$ gives finally (eq. 19):

$$\lim_{\beta \to \infty} \{\beta C_{q,\beta}^{2/q}(n)\} = \frac{2\left(\Gamma\left((q+1)/2\right)\right)^{2/q}}{\pi^{1/q}} \left(\frac{2}{\pi^2} \ln n + \frac{1}{2}\right) \qquad (23)$$

As discussed above, making $n = 1$, we are further led to:

$$C_{q,\beta}(1) = \left\langle |s-1|^q \right\rangle_\beta \approx B_{q,\beta}^{q/2} = c_q B_{2,\beta}^{q/2} \qquad (24)$$

Eq. 24 is an approximation all the better as $\beta$ is large. From eq. 21 and the exact values of the moments $\left\langle s^k \right\rangle_\beta$ [33], we calculate $B_{2,1} = 0.30268..$ and $B_{2,2} = 0.18375..$ which are reasonably close to the values, respectively of about 0.28 and 0.20, taken from figure 3 of [19]. The moments of the GG distribution calculated from eq. 8 lead to $B_{2,1} = 0.286$ and $B_{2,2} = 0.180$ respectively. The moments $\left\langle |s-1|^q \right\rangle_\beta$ approximated by those of the GG distribution are all the less accurate than $q$ is large. A more detailed explanation of the essentially linear variation of $B_{q,\beta}$ ($\beta = 1, 2$) with $q$ [19] is still needed.

### V.2.2. The simple logarithmic behavior
The functional dependence :

$$C_q(n) = \left(A_q \ln n + B_q\right)^{qH_q} \qquad (26)$$

with $H_q = H_0 = 1/2$, is named 'simple logarithmic behavior' in [19] since $C_q(n)$ has exponents similar to those of time series showing a simple scaling, $C_q(n) \propto n^{qH_0}$. Its name recalls that it has the same functional structure as the second-order correlation function (eq. 16), regardless of the value of $q$.

As shown in the previous section, a Gaussian distribution with a variance $\sigma^2 = A_2 \ln n + B_2$ leads to eq. 26 with:

$$A_q = \frac{2A_2\left(\Gamma\left((q+1)/2\right)\right)^{2/q}}{\pi^{1/q}}, \quad B_q = \frac{2B_2\left(\Gamma\left((q+1)/2\right)\right)^{2/q}}{\pi^{1/q}} \qquad (27)$$

The following argument suggests that the converse holds at least approximately in broad conditions when $A_q$ and $B_q$ are linear in $q$ as are almost perfectly the coefficients given by eq. 27. Let us consider a continuous and symmetric random variable $X$ whose even moments are:

$$\mu_{2j} = \langle x^{2j} \rangle = (\alpha j + \theta)^j \qquad (28)$$

with $\alpha, \theta > 0$. Then the expansion, when $t \to 0$, of its characteristic function $\phi(t) = \langle \exp(itx) \rangle$ reads:

$$\phi(t) = \sum_{j=0}^{\infty} (-1)^j \frac{(\alpha j + \theta)^j t^{2j}}{(2j)!} \qquad (29)$$

For $j \gg \theta/\alpha$ and large enough for the Stirling approximation of $j!$ to hold (it is already accurate to 1.4% for $j=6$), the $t^{2j}$ term reduces approximately to $\approx (-1)^j \dfrac{\left(\alpha e t^2 / 4^{1+1/4j}\right)^j}{j!}$ which is essentially the $t^{2j}$ term of the expansion of the characteristic function, $\phi(t) = \exp(-\alpha e t^2/4)$, of a Gaussian random variable $N(0, \alpha e/2)$. If eq. 28 reduces to $\theta^j$, then the distribution of $X$ becomes degenerate, $p(x) = \dfrac{1}{2}\left(\delta(x - \sqrt{\theta}) + \delta(x + \sqrt{\theta})\right)$.

## VI. $1/f$ SIGNALS

As summarized in the introduction, discrete 'time' series $\delta_n$ (eqs 1 and 12) associated with unfolded eigenvalues of random matrices from classical ensembles or with energy spectra of chaotic quantum systems exhibit $1/f^\alpha$ noise [15-27]. Before discussing our results on $1/f^\alpha$ noise in the β-Hermite ensemble, we consider two simple discrete time series which are interesting per se as they generate an exact $1/f^\alpha$ noise and enlighten the results of section V. Further, both are wide-sense stationary and display a simple logarithmic behavior [19,40].

### VI.1. A $1/f$ time series [40]

A $1/f$ time series analyzed by Greis and Greenside [40] was considered by Relaño et al. [19] as its variance is $\propto \ln n + cst$ (fig. 6 of [19]). It is defined as:

$$X(n) = \sum_{k=1}^{N/2} \frac{1}{\sqrt{k}} \cos\left(\frac{2\pi kn}{N} + \varphi_k\right) = \sum_{k=1}^{N/2} Y_{n,k} \qquad (30)$$

$n = 1,..,N$, where the $\varphi_k$'s are independent random variables uniformly distributed between 0 and $2\pi$. First the ensemble average $\langle X(n) \rangle = 0$ and:

$$s_N^2 = \langle X^2(n) \rangle = \frac{1}{2}\sum_{k=1}^{N/2} \frac{1}{k} \approx \frac{1}{2}\left[\ln\left(\frac{N}{2}\right) + \gamma + \frac{1}{N} + ..\right] \qquad (31)$$

are independent of $n$. Defining $r_N^3 = \sum_{k=1}^{N/2} \langle |Y_{n,k}|^3 \rangle$, it comes:

$$r_N^3 = \sum_{k=1}^{N/2} \frac{1}{k\sqrt{k}} \approx \frac{4\zeta(3/2)}{3\pi} \qquad (32)$$

and thus $\frac{r_N}{s_N} \propto \frac{1}{\sqrt{\ln N + cst}}$ from which the Lyapounov condition, $\lim_{N\to\infty}\left(\frac{r_N}{s_N}\right) = 0$, follows. The central limit theorem [42] shows then that $X(n)$ tends, whatever $n$, to a Gaussian $N(0, s_N^2)$. Computer simulations show that the Gaussian approximation is already valid for ensembles of realizations of some hundreds. The Lyapounov condition is no more valid for $X(m+n) - X(m)$ but its distribution tends to a Gaussian $N(0, C_2(n))$ when $n \ll N$ (appendix B). The ensemble average of the second-order correlation function:

$$C_2(n) = \langle (X(m+n) - X(m))^2 \rangle = \sum_{k=1}^{N/2} \frac{(1 - \cos(2\pi kn/N))}{k} =$$
$$= \ln(\sin^2(\pi n/N))/2 + \gamma + \ln N + Ci(n\pi) + O(1/N) \qquad (33)$$

gives :

$$\langle X(m+n) X(m) \rangle = -\ln(\sin^2(\pi n/N))/4 - \ln 2/2 + Ci(n\pi)/2 + O(1/N) \qquad (34)$$

for large $N$ and $1 \ll n$ from the cosine-integral correcting term, $Ci(x)$. The autocorrelation function decreases as $-\ln(n/N)/2$ for $n/N \ll 1$ and reaches its minimum for $n/N = 0.5$. When $1 \ll n \ll N$, $C_2(n)$ becomes:

$$C_2(n) = \ln n + \gamma + \ln \pi \qquad (35)$$

By contrast, when $n \sim N/2$, $C_2(n)$ rises as $\ln N$. From a least-squares fit of $C_2(n)$ by eq. 26, we obtain $A_2 = 0.956$ and $B_2 = 1.956$ for $N = 4096$ and $100 \leq n \leq 1000$. These values depend on the selected range of $n$, $B_2$ being naturally more sensitive ($\pm 0.07$) than $A_2$ ($\pm 0.01$) to that choice. For $N = 5.10^4$ and $1500 \leq n \leq 2500$, we get $A_2 = 0.995$ and $B_2 = 1.760$ in fair agreement with eq. 35 ($A_2 = 1$, $B_2 = 1.7219...$). The functional relation (eq. 26), $C_q(n) = (C_2(n))^{q/2}$, which is verified by the present time series $X(n)$, in the appropriate range of $n$, is then seen to be due again to an underlying Gaussian distribution with a variance $\propto \ln n + cst$.

### VI.2. Gaussian periodic $1/f^\alpha$ signals [43-45]

Antal et al. [43-45] investigated the extreme value statistics of periodic signals, either described as time signals or as 1D interfaces, displaying Gaussian fluctuations with $1/f^\alpha$ power spectra. They considered Gaussian periodic signals, $h(t) = h(t+T)$, of length $T$. Using the probability density functional of the 'height' $h(t)$, they derived the probability distributions of the amplitudes and of the phases of the coefficients $c_n$ of the Fourier expansion, $h(t) = \sum_{n=-N}^{N} c_n \exp(2i\pi n t/T)$ $(c_n^* = c_{-n}, c_0 = 0)$ whose real parts and imaginary parts are independent $N(0, \theta_n^2/4)$ Gaussian variables [44] with $\theta_n = 1/\sqrt{n^\alpha T^{1-\alpha}}$ ($\alpha > 0$ and the free parameter $\sigma$ of eq. 4 of [44] is taken here as 1). The height $h(t)$ is then a Gaussian $N\left(0, \sum_{n=1}^{N} \theta_n^2\right)$ whatever $t$. To emphasize its connection with the previous series, the model is conveniently formulated as follows:

$$h(t) = \sum_{k=1}^{N} s_k \cos\left(\frac{2\pi k t}{T} + \varphi_k\right) \qquad (36)$$

where the amplitudes $s_k = 2|c_k|$ and the phases $\varphi_k$ ($k = 1,..,N$) are mutually independent random variables, the latter being uniformly distributed between 0 and $2\pi$. The $s_k$'s have a Rayleigh distribution:

$$p(s_k) = s_k \exp(-s_k^2/2\theta_k^2)/\theta_k^2, \quad s_k \geq 0 \tag{37}$$

The cutoff introduced by $N$ prevents to resolve the time scale below $\tau = T/N$. Two time scales, the observation time $T$ and the microscopic time $\tau$, must thus be considered [44-45]. The 'roughness' is a time average over an entire period for a given realization [43-45]:

$$w_2 = \left\langle \left( h(t) - \langle h(t) \rangle_t \right)^2 \right\rangle_t = 2 \sum_{k=1}^{N} |c_k|^2 \tag{38}$$

where the second equality comes from the integrated power spectrum [44]. The roughness has the same meaning as the spectral average $\langle \delta_n^2 \rangle_m$ considered in section V. As $y_k = 2|c_k|^2/\theta_k^2$ is exponentially distributed, $p(y_k) = \exp(-y_k)$, it comes $\langle 2|c_k|^2 \rangle = \theta_k^2$ and a variance $\langle (2|c_k|^2 - \theta_k^2)^2 \rangle = \theta_k^4$. Thus:

$$\langle w_2 \rangle = \frac{1}{T^{1-\alpha}} \sum_{k=1}^{N} \frac{1}{k^\alpha} \qquad \mu_2 = \left\langle (w_2 - \langle w_2 \rangle)^2 \right\rangle = \frac{1}{T^{2(1-\alpha)}} \sum_{k=1}^{N} \frac{1}{k^{2\alpha}} \tag{39}$$

in agreement with eqs 17a and 17b of [44] that Antal et al. derived from the cumulant generating function. The autocorrelation function depends only on the time difference:

$$\langle h(t')h(t+t') \rangle = \frac{1}{T^{1-\alpha}} \sum_{k=1}^{N} \frac{\cos(2\pi kt/T)}{k^\alpha} \tag{40}$$

a result analogous to that of section VI.1. For $\alpha = 1$, a $\ln(T/\tau)$ dependence is thus found for the average roughness $\langle w_2 \rangle$ [44]. When $N = T/\tau \to \infty$, the autocorrelation function is then (eq. 34) $-\ln(\sin^2(\pi t/T))/2 - \ln(2) + Ci(2\pi t/\tau) + O(\tau/T)$. It varies logarithmically, as $-\ln(t/T) + cst$, when $\tau/T \ll t/T \ll 1$. Arguments similar to those of the previous sections might be used to show that the present Gaussian series exhibits a simple logarithmic behaviour when $\alpha = 1$ and $\tau/T \ll t/T \ll 1$.

To summarize, logarithmic dependences are ubiquitous in the fluctuation characteristics of the three discrete series discussed above that display $1/f$ noise. Their mean-square fluctuations, $\langle \delta_n^2 \rangle$ when $n$ is not small as compared to $N$, $\langle X^2(n) \rangle$ and $\langle w_2 \rangle$ vary logarithmically with the size $N$ (or $T$). They show a logarithmic dependence of the autocorrelation function, with a negative slope, when $1 \ll y \ll Y$ ($y = n, t$, $Y = N, T$). In all cases, limiting Gaussian distributions explain the observed 'simple logarithmic' behaviors with variances depending linearly on $\ln y$, when $y$ is small enough, but not too small, as compared to the overall size of the considered series. This is a consequence of the presence of a factor $1/k$ in the $k$ th term of the sum defining the variances. We notice that wide-sense stationary time series that display limited $1/f$ noise, that is a noise which cannot be distinguished experimentally from exact $1/f$ noise over a wide frequency range $(1/2\pi\tau_2, 1/2\pi\tau_1)$, have autocorrelation functions which vary as $-\ln t$ for $\tau_1 < t < \tau_2$ [46-47]. The Wiener-Khintchine theorem can indeed be used to write the autocorrelation function as $K(t) = c \int_{1/\tau_2}^{1/\tau_1} df \cos(2\pi ft)/2\pi f$ [46] whose discrete counterpart appears above.

The $\delta_n$ statistics of the Gaussian ensembles ($\beta = 0, 1, 2, 4$ for the GDE, GOE, GUE, GSE respectively) exhibit a $1/f^2$ for $\beta = 0$ and $1/f$ noise for $\beta \geq 1$ [15-27]. A direct characterization of the $1/f^\alpha$ noise of the $\delta_n$ time series of the $\beta$-Hermite ensemble is described below when $\beta$ spans the interval $(0,\sim 32)$ to follow the concomitant evolution of $\alpha$ from 2 to 1.

### VI.3. Power spectrum of $\delta_n$ for the $\beta$-Hermite ensemble

The Fourier analysis was performed for the $\delta_n$ time series via the average power spectrum of the signal defined as :

$$P(k) = \frac{1}{N} \left\langle \left| \sum_{j=1}^{N} \delta(j) \exp\left(-i\frac{2\pi jk}{N}\right) \right|^2 \right\rangle, \quad k = 1,..,N \qquad (41)$$

It was calculated from 5000 $\beta$-H matrices for $N = 513$. The variation with $\beta$ of the slope, $-\alpha$, of the linear variation of $P(k)$ over two decades is shown in figure 7. The exponent $\alpha$ decreases smoothly from 2 to 1 when $\beta$ increases from 0 to $\beta \approx 1$, with $\alpha = 1.5$ for $\beta = \beta_m \approx 0.01$. As the wavelet transform constitutes a more efficient method to obtain the exponent $\alpha$ for the $\delta_n$ statistic [26], we used that method to follow $\alpha(\beta)$ as a function of matrix size.

## VI.4. Wavelet analysis

As signal processing methods based on wavelets are now widespread, we refer the reader to classical books [48-49] and we sketch only briefly the method we used [50-54]. A wide-sense stationary time series $X(t)$ can be formally written as [51-52]:

$$X(t) \propto \sum_{j,k=-\infty}^{+\infty} d_X(j,k)\psi_{j,k}(t) \qquad (42)$$

where the $\psi_{j,k}(t)$'s are obtained from a mother wavelet $\psi(t)$ by dyadic dilations and integer translations:

$$\psi_{j,k}(t) = \frac{1}{2^{j/2}} \psi\left(\frac{t}{2^j} - k\right) \qquad (43)$$

The mother wavelets considered here have $M(\geq 2)$ zero moments $\int_{\mathbb{R}} t^m \psi(t)dt = 0 \; (m=0,..,M-1)$. The coefficients $d_X(j,k)$ of the discrete wavelet transform, quantify frequency details of $X$ at scale $j$ and at location $k$. When the scale $j$ is large, the coefficient $d_X(j,k)$ captures low-frequency or coarse-scale behavior of $X(t)$. Conversely, the coefficient $d_X(j,k)$ characterizes the high-frequency or fine-scale details of $X(t)$ at small scales $j$ [52]. These coefficients are defined by:

$$d_X(j,k) = \frac{1}{2^{j/2}} \int_{-\infty}^{+\infty} X(t) \psi\left(\frac{t}{2^j} - k\right) dt \qquad (44)$$

If $n_j$ is the number of coefficients at scale $j$, the variance $v_X(j)$ of $d_X(j,k)$, $v_X(j)$, also denoted as the mean energy of the wavelet coefficients at scale $j$, can be estimated from [53]:

$$v_X(j) = \left(\sum_{k=1}^{n_j} d_X(j,k)^2\right) \Big/ n_j \qquad (45)$$

as the mean of $d_X(j,k)$ is zero by construction [53]. The wavelet energy spectrum, defined as the set of variances $v_X(j)$, is related to the power spectrum of the time series $X(t)$ [50-54]. The wavelet energy spectrum summarises the spectrum information using just one value per frequency band and is of interest in particular when the power spectrum is relatively featureless in each band [51]. When the time series displays $1/f^\alpha$ noise, then in rather mild conditions [51-54]:

$$v_X(j) \approx cst\, 2^{j+1} \int_{2^{-(j+1)}}^{2^{-j}} \frac{df}{f^\alpha} \qquad (46)$$

which gives a linear relationship between $\ln_2(v_X(j))$ and $j$ with a slope $\alpha$.

$N \times N$ $\beta$-Hermite matrices with $N = 2^p + 1$, with $p$ varying from 9 to 15, were unfolded to obtain $2^{p-1}$ spacings. The wavelet analysis of the $\delta_n$ series was performed over $p-1$ scales with a number of coefficients decreasing from $2^{p-2}$ for the finest scale to 1 for the coarser using the Wavelab software (version 850) [55]. The first set of $2^{p-2}$ noisy coefficients and the last coefficient were discarded and the linear regression described above was performed from the ensemble average of the second moments of the wavelet coefficients of the $p-3$ internal scales. Then, as the number of coefficients at scale $j$, $n_j \propto 2^{-j}$, is here $\propto 2^p$, $p$ is equal to $-j$ except for an irrelevant shift independent of $j$. Linear relationships are then expected to hold between $\ln_2(v_\delta(p))$ and $p$ with a slope $-\alpha$ as convincingly shown by figure 8. Different wavelets were used and seen to show the same bevavior of $\alpha(\beta)$ as those found with Daubechies wavelets of different indices (figure 9) which have compact support in the time domain and a well-localized support in the frequency domain [48-49]. Figure 8 shows first that that the $\delta_n$ series is characterized by a $1/f$ noise for any $\beta \geq 1$ in agreement with the results found for the three classical Gaussian ensembles $(\beta = 1, 2, 4)$ [15-27]. When $\beta$ decreases from 1 to 0, the noise evolves from $1/f$ at large $\beta$ to $1/f^2$ when $\beta$ is close to zero. An homogeneous evolution would exhibit a single intermediate $1/f^\alpha$ noise with $1 < \alpha < 2$ at all scales but figure 8b shows that it is heterogeneous with a $\sim 1/f^2$ noise at the finest scales and a $\sim 1/f$ noise at the coarsest ones. The analysis of the transition was nevertheless performed from the slopes of linear fits to the various curves. Therefore, a value of $\alpha$ intermediate between 1 and 2 is a convenient effective value but it does not necessarily mean that it results from a $1/f^\alpha$ noise at all scales. For instance, the slope obtained for $\beta = 1/128$ and $N = 32769$ from a linear fit of all points is $\alpha = 1.50$ which is in that case the average of the slopes fitted from the zones $3 \leq p \leq 8$ and $9 \leq p \leq 14$ which are $\alpha_1 = 1.19$ ($\sim 1/f$) and $\alpha_2 = 1.83$ ($\sim 1/f^2$) respectively. Simulations and wavelet analyses were performed for $\beta = 2^{-m}$ with $m = -3$ to 13. For clarity, only some of the obtained results are shown on figure 8b. When $\beta$ decreases in the region where $\alpha$ increases rapidly, the range of scales in which the $\sim 1/f^2$ noise predominates increases (figure 8b).

The analysis of the effect of the matrix size on the $\alpha(\beta)$ curves was performed with a Daubechies wavelet of index 10. All the $\alpha(\beta)$ curves are very well described by (figure 10a):

$$\alpha(\beta) = 1.5 - 0.5 \times erf\left(\log(\beta/\beta_m)/\sigma_\alpha \sqrt{2}\right) \qquad (52)$$

The least-squares fitted parameters $\beta_m$ and $\sigma_\alpha$ are given in table 2. Figure 10b shows that the four curves $\alpha\left(\log(\beta/\beta_m)/\sigma_\alpha\right)$ merge indeed together and suggests that a unique growth mechanism of the $\sim 1/f^2$ fine-scales operates whatever the matrix size when it is large enough. The parameter $\beta_m$ decreases rather slowly with $N$ as $\beta_m \approx 1.1/N^{0.47}$ while the apparent small increase $\sigma_\alpha$ may not be significant. In any case, the curves $\alpha(\beta)$ are shifted downwards without becoming steeper when $N$ increases. We conclude that the $1/f^2$ behaviour occurs only asymptotically at $\beta = 0$ while it is the $1/f$ behaviour which is the rule for $\beta > 0$.

## VII. CONCLUSION

The $\beta$-Hermite ensemble makes it possible to investigate efficiently the $\delta_n$ statistic, the fluctuation of the $n$ th unfolded eigenvalue, where $n$ plays the role of a discrete time, for any $\beta > 0$ and thus to extend previous results found for the three classical Gaussian ensembles $(\beta = 1, 2, 4)$. The spacing variance varies as $2\ln n/\beta\pi^2 + cst(\beta)$ for any $\beta > 0$, when $1 \ll n \ll N$, where $\beta \times cst(\beta) \to 1/2$ when $\beta \to \infty$. For large values of $n$ ($\sim N/4$), the variance $\langle \delta_n^2 \rangle$ increases as $\ln N$. The autocorrelation function depends on $\ln n$ whatever $\beta$ when $1 \ll n \ll N$ with an overall level rising as $\ln N$. The simple logarithmic behavior shown by the higher-order moments of $\delta_n$ is accounted for by Gaussian distributions whose variances depend linearly on $\ln n$. Analogous results are found for two known time series constructed to exhibit $1/f$ noise. The $1/f^\alpha$ noise displayed by the $\delta_n$ series is characterized by wavelet analysis for the $\beta$-Hermite ensemble both as a function of $\beta$ and of matrix size $N$. When $\beta$ decreases from 1 to 0 for a given and large enough matrix size, the evolution from a $1/f$ noise to a $1/f^2$ noise does not take place homogeneously through an intermediate $1/f^\alpha$ noise at all scales but heterogeneously through a mixture of a $\sim 1/f^2$ noise at the finest scales and of a $\sim 1/f$ noise at the coarsest ones. The $\sim 1/f^2$ range grows when $\beta$ decreases down to 0. Asymptotically, a $1/f^2$ noise is found for $\beta = 0$ while a $1/f$ noise is the rule for $\beta > 0$. The $1/f$ behaviour is related to the small amplitude normal modes which are essentially plane waves in the limit of large matrices as shown by Andersen et al. [34] with a mean square amplitude proportional to $1/k$ for the $k$ th mode (see too [30]).

**APPENDIX A : THE SPACING VARIANCE AND THE AUTOCORRELATION FUNCTION DERIVED FROM EQ. 15**

For large $N$, the multivariate Gaussian approximation of the joint distribution of eigenvalues [30, 34] and the resulting distribution of unfolded eigenvalues in the central part of the spectrum [34] allow calculation of the moments of the distribution of the spacing $S_n = S_n(0) = \sum_{k=1}^{n} s_k$ between successive unfolded eigenvalues (eq. 12). With the condition that $\langle s_k \rangle = 1$, eq.15 gives:

$$S_n = n + \frac{\sqrt{2N}}{\pi} \sum_{j=1}^{N} \alpha_j \left( U_{j-1}\left(\frac{\pi(n+1)}{2N}\right) - U_{j-1}\left(\frac{\pi}{2N}\right) \right) \quad \text{(A-1)}$$

The $U_{j-1}(x)$'s are Chebyshev polynomials of the second kind and the $\alpha_j$'s are independent $N\left(0, \frac{1}{j\beta N}\right)$ Gaussians [34]. The variance $\sigma_\beta^2(n)$ is then calculated from :

$$\sigma_\beta^2(n) = \langle S_n^2 \rangle - n^2 = \frac{2}{\beta \pi^2} \sum_{j=1}^{N} \frac{1}{j} \left( U_{j-1}\left(\frac{\pi(n+1)}{2N}\right) - U_{j-1}\left(\frac{\pi}{2N}\right) \right)^2 \quad \text{(A-2)}$$

It involves ensemble averages of terms of the form:

$$\langle T_N^2(a,b) \rangle = \sum_{j=1}^{N} \frac{1}{j} \left( U_{j-1}\left(\frac{\pi a}{2N}\right) - U_{j-1}\left(\frac{\pi b}{2N}\right) \right)^2 \quad \text{(A-3)}$$

The Chebyshev polynomial $U_{j-1}(y)$ can be written as $U_{j-1}(\cos(x)) = \sin(jx)/\sin(x)$, when $|y| < 1$, with $y_a = \cos(x_a) = \pi a/2N$ and $y_b = \cos(x_b) = \pi b/2N$ (when $y \geq 1$, $U_{j-1}(y)$ is given instead by $U_{j-1}(\cosh(x)) = \sinh(jx)/\sinh(x)$). The series [41]:

$$\sum_{j=1}^{\infty} \frac{\cos(jx)}{j} = -\frac{1}{2}\ln\left(4\sin^2\left(\frac{x}{2}\right)\right) \quad (0 < x < 2\pi) \qquad \text{(A-4)}$$

, with the eventual addition of a cosine integral term $Ci(Nx)$ to correct for the effect of a finite summation from 1 to $N$, and the classical series $\sum_{j=1}^{N} \frac{1}{j} = \gamma + \ln N + O\left(\frac{1}{N}\right)$, allows us to calculate

$$P_N(a,b) = \left\langle \sum_{j=1}^{N} \frac{1}{j}\left(U_{j-1}\left(\frac{\pi a}{2N}\right)U_{j-1}\left(\frac{\pi b}{2N}\right)\right)\right\rangle \quad \text{for } a \text{ and } b \text{ fixed, } N \text{ large and } a \neq b.$$ 

For $1 \ll a, b \ll N$, $P_N(a,b)$ becomes:

$$P_N(a,b) = \frac{\ln N}{2} - \frac{\ln|a-b|}{2} + \ln(2) - \frac{\ln \pi}{2} + O\left(\frac{1}{N}\right) \qquad \text{(A-5)}$$

and:

$$P_N(a,a) = \frac{1}{2}(\ln N + \gamma + \ln 2) + O\left(\frac{1}{N}\right) \qquad \text{(A-6)}$$

Finally, from eqs A-2, A-5 and A-6, we obtain:

$$\sigma_\beta^2(n) = \frac{2}{\beta \pi^2}\left(\ln n + \gamma + \ln(\pi/2)\right) \qquad \text{(A-7)}$$

The number variance $\Sigma_\beta^2(L)$, which was calculated by Andersen et al. [34] from the same starting point (eq. 15), is such that $\sigma_\beta^2(n) \equiv \Sigma_\beta^2(n)$ as expected from eq. 14, except for the constant term $-1/6$ solely related to the classical Gaussian ensembles. The lower order terms are only rough approximations of the exact terms in contrast to the logarithmic terms which are correctly obtained by that approximation method [34].

Similar calculations were performed from eq. 15 for the autocorrelation function $K_\beta(n) = \left\langle (\delta_{m+n}\delta_m)_m \right\rangle$. As the lower order terms are not obtained by that method, we focus solely on the leading term which comes from:

$$\left\langle \delta_{m+n}\delta_m \right\rangle_m = \frac{2N}{\pi^2}\left\langle T_N(m+n,1)T_N(m,1)\right\rangle_m \qquad \text{(A-8)}$$

As the ensemble average cancels products like $\alpha_j \alpha_k$ with $j \neq k$, because of the independence of the normal variables $\alpha_j$ and $\alpha_k$, it suffices to consider those which involve $\alpha_j^2$. The only products which gives rise to an $n$ dependent term are of the form $U_{j-1}\left(\dfrac{\pi(m+n+1)}{2N}\right)U_{j-1}\left(\dfrac{\pi(m+1)}{2N}\right)$. Interverting the two summations, the first performed on a given realization (eq. A-8) with $m$ running between 1 and $M$ and the second over realizations, we obtain, when $N$ increases, the following contribution for given $m$ and $1\ll n\ll N$:

$$K_\beta(n) = \left\langle (\delta_{m+n}\delta_m)_m \right\rangle = -\dfrac{1}{\beta\pi^2}\ln n + K_\beta(1) \qquad \text{(A-9)}$$

when $M$ is fixed or when $M$ increases as $M = \alpha N$ with $\alpha \ll 2/\pi$ and where $K_\beta(1)$ is expected to depend on $N$. The slope $-1/\beta\pi^2$ agrees with numerical simulations for $\sim 0.005 < n/N < \sim 0.1$ (figure 6).

## APPENDIX B : CUMULANT FUNCTION OF $X(m+n)-X(m)$ (Eq. 30)

Here $m$ is fixed and averages are ensemble averages. The characteristic function of $X(m+n)-X(m)$ is:

$$\begin{aligned}\Phi_{m,n}(t) &= \left\langle \exp\left(it\left(X(m+n)-X(m)\right)\right)\right\rangle \\ &= \prod_{k=1}^{N/2}\left\langle \exp\left[2it\dfrac{\sin(\pi kn/N)}{\sqrt{k}}\sin\left(\phi_k + 2\pi k(m+n/2)/N\right)\right]\right\rangle_{\phi_k}\end{aligned} \qquad \text{(B-1)}$$

The cumulant function, which is independent of $m$, is then:

$$\ln\Phi_{m,n}(t) = \sum_{k=1}^{N/2}\ln\left(\dfrac{1}{2\pi}\int_0^{2\pi}\cos\left(2t\dfrac{\sin(\pi kn/N)}{\sqrt{k}}\sin\left(\phi_k + 2\pi k(m+n/2)/N\right)\right)d\phi_k\right) \qquad \text{(B-2)}$$

Its expansion reads:

$$\ln\Phi_{m,n}(t) = \sum_{k=1}^{N/2}\left(-\dfrac{\sin^2(\pi kn/N)}{k}t^2 - \dfrac{\sin^4(\pi kn/N)}{4k^2}t^4 - \dfrac{\sin^6(\pi kn/N)}{9k^3}t^6 + O(t^8)\right) \qquad \text{(B-3)}$$

that is:

$$\ln\Phi_{m,n}(t) = -\dfrac{C_2(n)}{2}t^2 - \dfrac{\pi^2 n}{16N}t^4 + .. \qquad \text{(B-4)}$$

where higher-order terms in $n/N$ are $O\left((n/N)^2\right)$. When $n \ll N$, the cumulant function reduces to $\ln \Phi_{m,n}(t) = -\dfrac{C_2(n)}{2}t^2$ and the distribution of $X(m+n) - X(m)$ tends thus to a Gaussian distribution $N(0, C_2(n))$.

Table 1 : A comparison of the values of $B_{2,\beta}$, calculated from eqs 13 and 14 and from eq. 16 for the three classical Gaussian ensembles, to the exact NNS variances $\sigma_\beta^2(1)$ [33]. The two shape parameters of the GG distribution (eq. 8), $\omega_1 = \beta$ and $\omega_2$, are taken from table 1 of [32]. Approximate NNS variances are then calculated from eq. 17.

| $\beta$ | $\omega_2$ [32] | $\sigma_\beta^2(1)$ GG (eq. 17) | $\sigma_\beta^2(1)$ exact [33] | $B_{2,\beta}$ (eqs 13-14,16) |
|---|---|---|---|---|
| 1 | 1.886 | 0.285567 | 0.28553065.. | 0.27537580.. |
| 2 | 1.973 | 0.180058 | 0.17999387.. | 0.17935457.. |
| 4 | 2.007 | 0.104149 | 0.10409842.. | 0.10395919.. |

Table 2 : The $N$ dependence of the parameters $\beta_m$, defined as $\alpha(\beta_m) = 1.5$, and $\sigma_\alpha$, which are obtained by least-squares fitting the $\alpha_N(\beta)$ curves of figure 10a by eq. 52.

| $N$ | $\log(\beta_m)$ | $\beta_m \cdot 10^2$ | $\sigma_\alpha$ |
|---|---|---|---|
| 513 | -1.23 (0.01) | 5.90 (0.14) | 0.87 (0.02) |
| 2049 | -1.52 (0.02) | 3.02 (0.14) | 0.99 (0.03) |
| 8193 | -1.79 (0.04) | 1.63 (0.15) | 1.01 (0.06) |
| 32769 | -2.07 (0.04) | 0.85 (0.08) | 1.18 (0.06) |

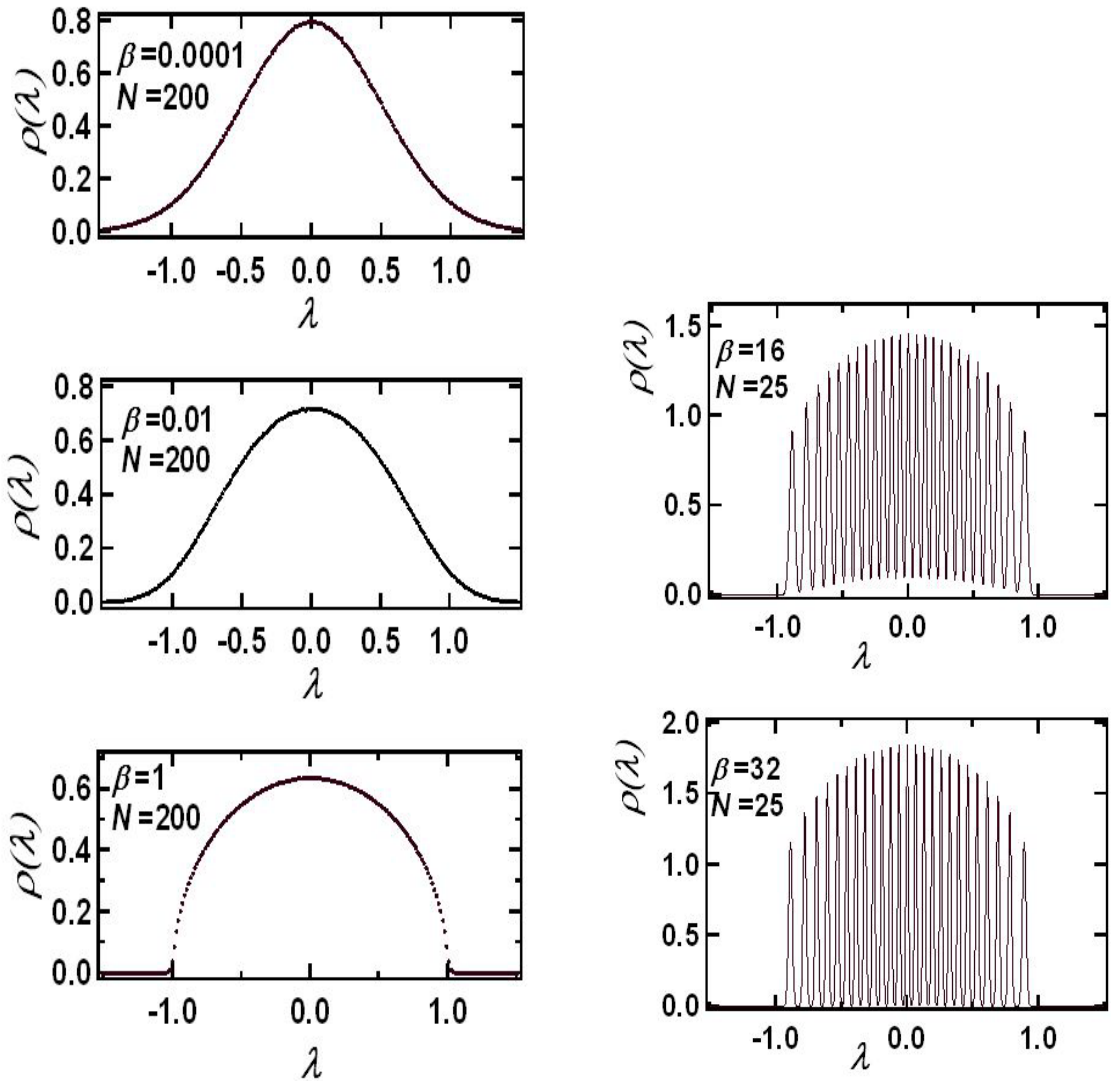

Figure 1 : Simulated eigenvalue densities $\rho(\lambda)$ of $N \times N$ $\beta$-Hermite matrices as a function of $\beta \, (\beta \leq 1)$, $\beta = 0.0001$, $0.01$, $1$, for fixed $N = 200$ and for $\beta > 1$, $\beta = 16$ and $32$ with $N = 25$. In all cases, $\langle \lambda^2 \rangle = \dfrac{1}{4}$.

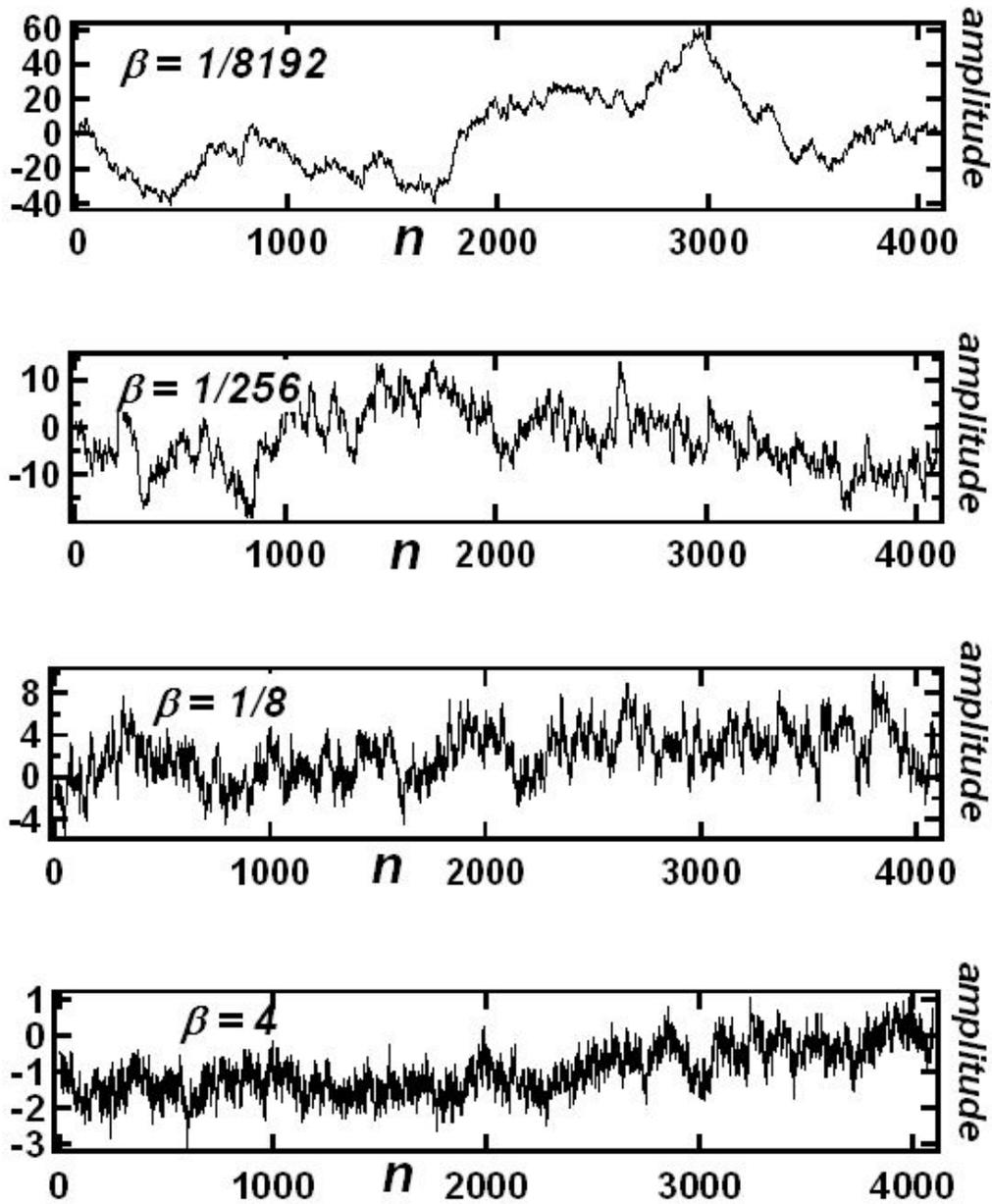

Figure 2 : Four realizations of the process, $\delta_n = \sum_{i=1}^{n} s_i - n$, from the unfolded eigenvalues of $N \times N$ $\beta$-Hermite matrices for the indicated values of $\beta$ and for $N = 4097$.

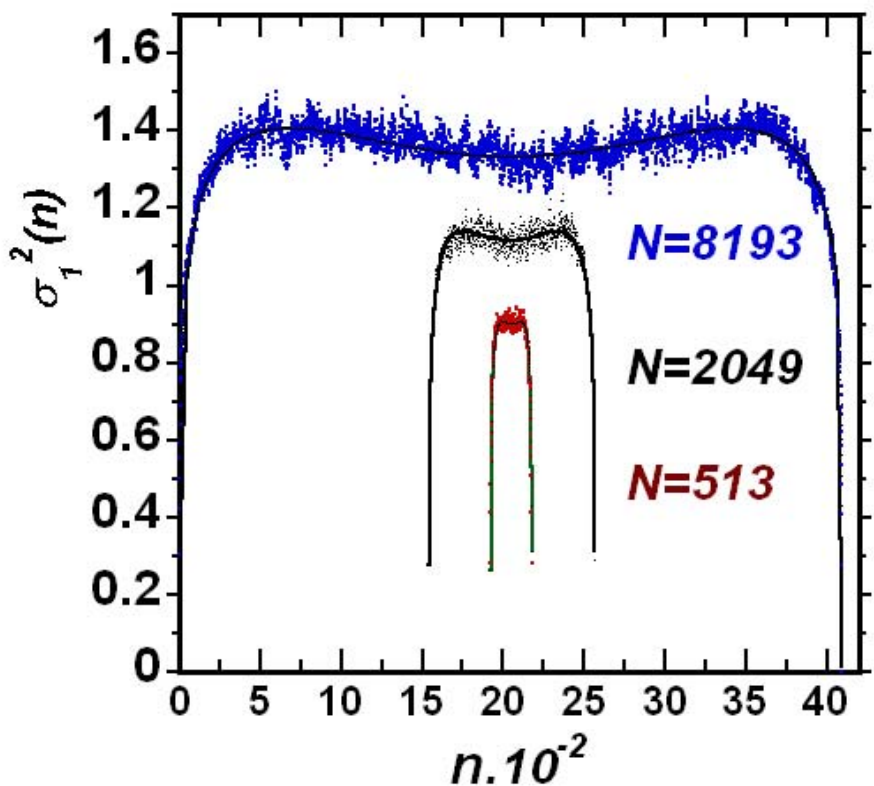

Figure 3 : Spacing variance, $\sigma_1^2(n)$, as a function of $n$ for $\beta = 1$, as calculated from the full sets of unfolded eigenvalues of $N \times N$ $\beta$-Hermite matrices for the indicated values of $N$. Solid lines are tentative fits of the simulated data by $a_1 \ln\left(\sin^2\left(2\pi n/N\right)\right) + b_{1,N} + c_{1,N}\left(1 - 4n/N\right)^2$.

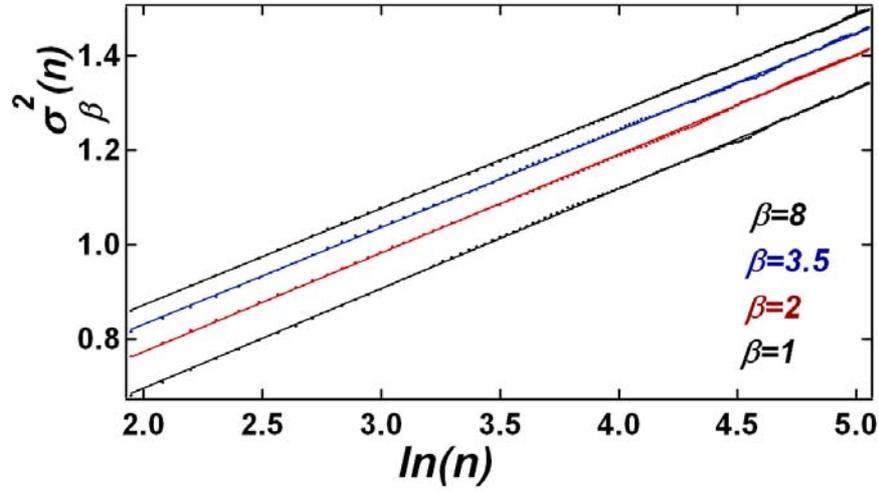

Figure 4 : The spacing variance $\sigma_\beta^2(n)$, calculated from Monte Carlo simulations and least-squares fitted by $a_\beta \ln n + b_\beta$, as a function of $\ln n$ $(7 \leq n \leq 148)$ for different values of $\beta$ ($\beta$ increases from bottom to top).

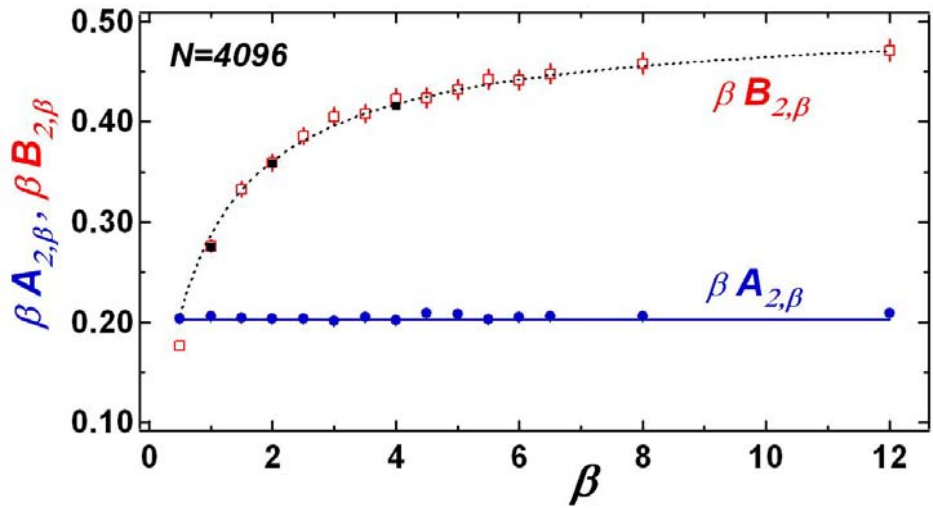

Figure 5 : The parameters $\beta A_{2,\beta}$ (solid circles) and $\beta B_{2,\beta}$ (empty squares) of the spacing variance $\sigma_\beta^2(n)$ (eq. 16), calculated from Monte Carlo simulations as a function of $\beta$. The solid line is $a_2 = \beta A_{2,\beta} = 2/\pi^2$, the dotted curve is $\beta \sigma_\beta^2(1)$ calculated from eq.17 with $\omega$ given by eq. 8. The solid squares correspond to the theoretical values for $\beta = 1, 2, 4$.

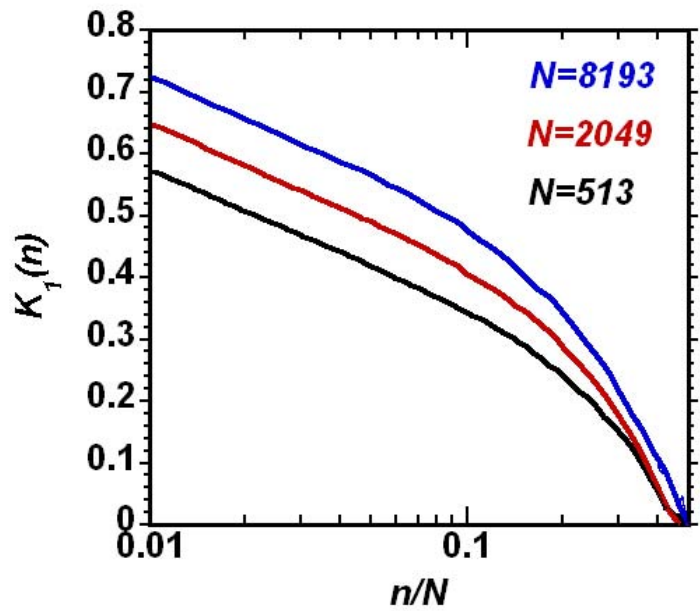

Figure 6 : Autocorrelation function, $K_\beta(n) = \langle (\delta_{m+n}\delta_m)_m \rangle$, as a function of $\ln n$ for $\beta=1$ and the indicated values of $N$.

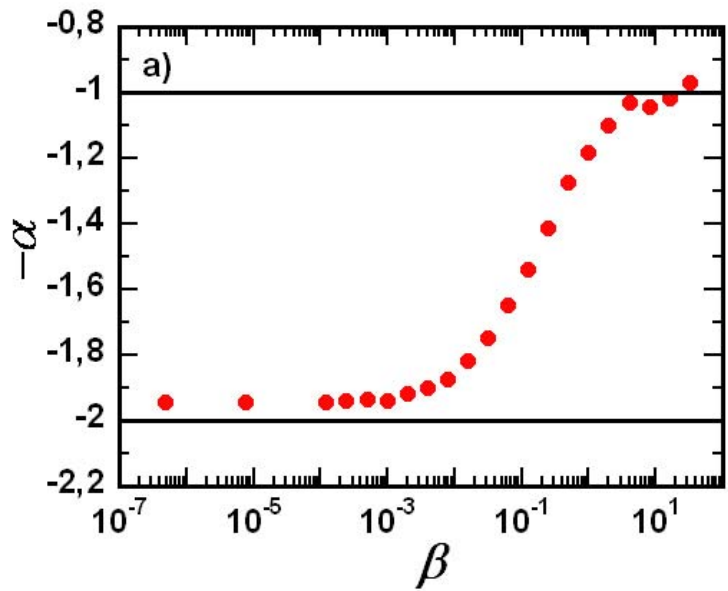

Figure 7: The variation of the exponent $\alpha$ as a function of $\beta$ as deduced from the power spectra of the $\delta_n$ series obtained from $513\times 513$ $\beta$-Hermite matrices.

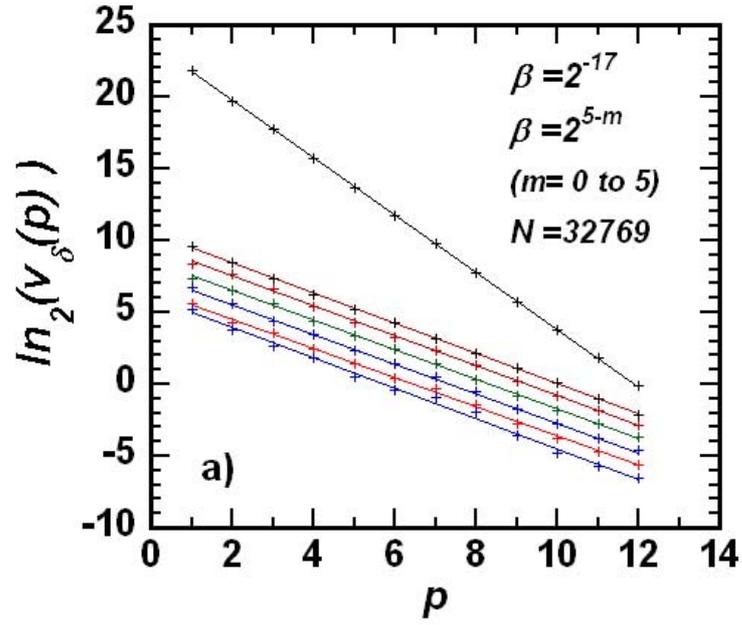

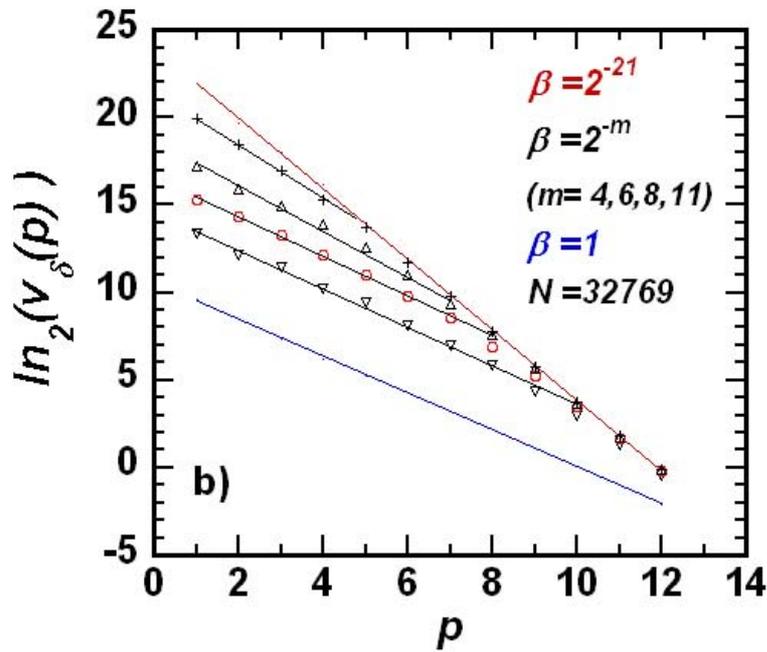

Figure 8: The log-log plot of the variance $v_\delta(p)$ of the wavelet coefficients of the $\delta_n$ series as a function of $p$ for different values of $\beta$ and of $N$; $\beta$ decreases as indicated from bottom to top a) from $\beta = 2^5$ to $\beta = 2^{-17}$ b) from $\beta = 1$ to $\beta = 2^{-21}$.

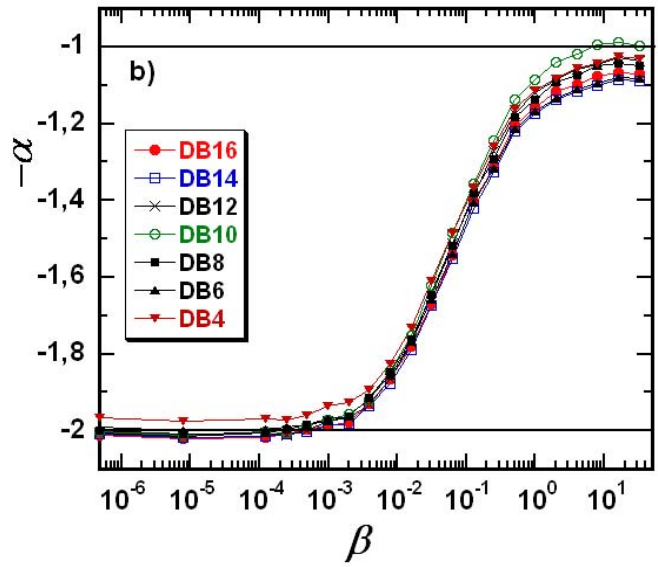

Figure 9 : The variation of the exponent $\alpha$ as a function of $\beta$ from the analysis of $\delta_n$ series with different Daubechies wavelets ( $N = 513$ ).

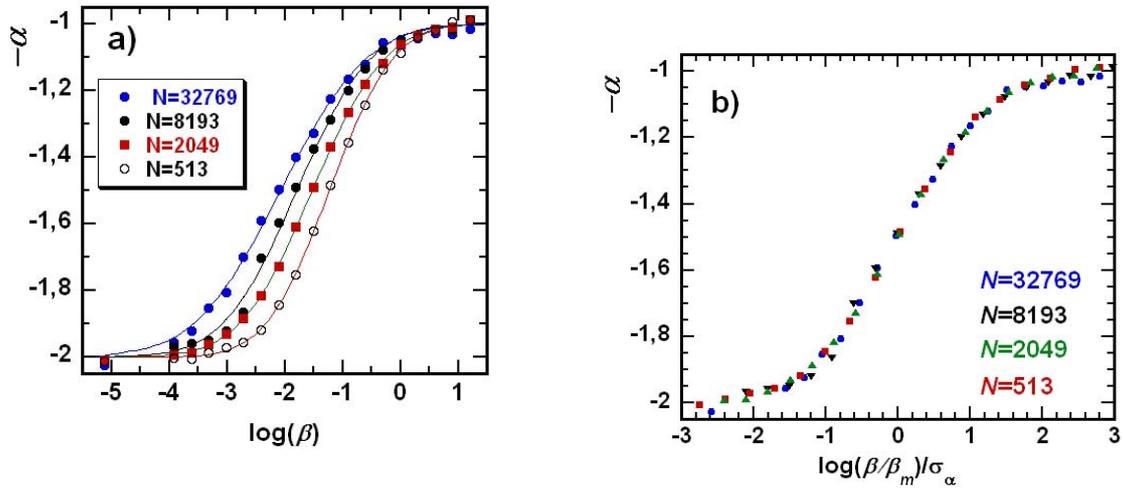

Figure 10: a) The variation of the exponent $\alpha$, as a function of $\beta$ and $N$, from the analysis of $\delta_n$ series with the Daubechies wavelet of index 10. The solid lines are the best fits of $\alpha(\beta)$ by eq. 52. b) Rescaling of the curves $\alpha(\beta)$ shown in a) with the parameters given in table 2.